\documentclass[usenatbib,useAMS,letterpaper]{mnras}
\usepackage{float}
\usepackage{graphicx}
\usepackage{fixltx2e}
\usepackage{epstopdf}
\usepackage{ textcomp }

\usepackage{ctable}
\usepackage{url}
\usepackage{times}
\usepackage{amsmath}
\usepackage{amssymb}
\usepackage{xspace}
\usepackage{mathrsfs} 
\usepackage{tabularx}
\usepackage{fixltx2e}
\usepackage{color}
\usepackage{placeins}

\newcommand{\acknowledgments}{\begin{small}\section*{Acknowledgements}\end{small}}

\newcommand\sref[1]{\hyperref[#1]{\S~\ref*{#1}}}
\newcommand\fref[1]{\hyperref[#1]{Fig.~\ref*{#1}}}
\newcommand\Eqref[1]{equation~(\hyperref[#1]{\ref*{#1}})}
\newcommand\tref[1]{\hyperref[#1]{Table~\ref*{#1}}}
\newcommand\aref[1]{\hyperref[#1]{Appendix~\ref*{#1}}}

\definecolor{green(pigment)}{rgb}{0.0, 0.65, 0.31}

\newcommand{\ksu}[1]{\textcolor{black}{#1}}

\title[Stellar feedback \& magnetic fields]{Stellar feedback strongly alters the amplification and morphology of galactic magnetic fields}
\author[Su, Hayward, Hopkins et al.]{
\parbox[t]{\textwidth}{
Kung-Yi Su$^{1}$\thanks{E-mail: ksu@caltech.edu}, Christopher C. Hayward$^{2,3}$, Philip F. Hopkins$^{1}$,  Eliot Quataert$^4$, Claude-Andr\'e Faucher-Gigu\`ere$^5$, Du\v san Kere\v s$^6$
}
\vspace*{6pt} \\
$^1$TAPIR 350-17, California Institute of Technology, 1200 E. California Boulevard, Pasadena, CA 91125, USA\\
$^2$Center for Computational Astrophysics, Flatiron Institute, 162 Fifth Avenue, New York, NY 10010, USA\\
$^3$Harvard-Smithsonian Center for Astrophysics, 60 Garden Street, Cambridge, MA 02138, USA\\
$^4$Department of Astronomy and Theoretical Astrophysics Center, University of California Berkeley, Berkeley, CA 94720, USA\\
$^5$Department of Physics and Astronomy and CIERA, Northwestern University, 2145 Sheridan Road, Evanston, IL 60208, USA\\
$^6$Department of Physics, Center for Astrophysics and Space Sciences, University of California at San Diego, 9500 Gilman Drive, La Jolla, CA 92093, USA\\
}
\usepackage[all]{hypcap}

\begin{document}
\long\def\/*#1*/{}
\date{Submitted to MNRAS}

\pagerange{\pageref{firstpage}--\pageref{lastpage}} \pubyear{2017}

\maketitle

\label{firstpage}

\begin{abstract}
Using high-resolution magnetohydrodynamic simulations of idealized, non-cosmological galaxies, we investigate how cooling, star formation, and stellar feedback affect galactic magnetic fields.
We find that the amplification histories, saturation values, and morphologies of the magnetic fields vary considerably depending on the baryonic physics employed, primarily because of differences in the gas density distribution.
In particular, adiabatic runs and runs with a sub-grid (effective equation of state) stellar feedback model yield lower saturation values and morphologies that exhibit greater large-scale order compared with runs that adopt explicit stellar feedback and runs with cooling and star formation but no feedback. The discrepancies mostly lie in gas denser than the galactic average, which requires cooling and explicit fragmentation to capture.
Independent of the baryonic physics included, the magnetic field strength scales with gas density as $B\propto n^{2/3}$,
suggesting isotropic flux freezing or equipartition between the magnetic and gravitational energies during the field amplification.
We conclude that accurate treatments of cooling, star formation, and stellar feedback are crucial for obtaining the correct magnetic field strength and morphology in dense gas, which, in turn, is essential for properly modeling other physical processes that depend on the magnetic field, such as cosmic ray feedback.
\end{abstract}


\section{Introduction} \label{S:intro}
Magnetic fields may play a role in galaxy evolution because the magnetic pressure can reach approximate equipartition with the turbulent or/and thermal pressure \citep{1996ARA&A..34..155B,2009ASTRA...5...43B}. Idealized simulations of isolated galaxies with magnetic fields suggest that magnetic fields can provide extra support in dense clouds and suppress star formation \citep{2009ApJ...696...96W,2012MNRAS.422.2152B,2013MNRAS.432..176P}, and magnetic acceleration has been suggested as a mechanism to drive outflows \citep{1982MNRAS.199..883B}. Moreover, magnetic fields can also suppress fluid mixing instabilities, including the Rayleigh-Taylor and Kelvin-Helmholtz instabilities \citep{1995ApJ...453..332J}, and therefore also largely suppress the `cloud shredding' process \citep{2015MNRAS.449....2M}. 

In \citet{2016arXiv160705274S}, we performed high-resolution magnetohydrodynamic (MHD) simulations incorporating multi-channel explicit stellar feedback, including both cosmological `zooms' and idealized isolated galaxies to investigate the effect of magnetic fields and other `microphysics' on galaxy formation. We found that the presence of a magnetic field had at most $\sim 10$-per cent-level effects on global galaxy properties such as the star formation history (SFH) because the turbulence in the simulated galaxies' ISM is super-Alfv{\'e}nic; thus, the magnetic field is dynamically unimportant compared with turbulence.

However, our previous results do \emph{not} necessarily imply that magnetic fields can be neglected in galaxy formation simulations. In particular, the simulations presented in \citet{2016arXiv160705274S} did not include cosmic rays, which may be an important driver of galactic outflows \citep{Uhlig2012,Hanasz2013,Booth2013,Salem2014a}. Because cosmic rays are coupled to magnetic fields, to treat cosmic ray transport correctly, one must first accurately determine the magnetic field \citep[e.g.][]{Pakmor2016}, which requires correctly capturing various potentially important amplification mechanisms, including flux freezing compression and the turbulent and $\alpha-\Omega$ dynamos \citep[e.g.][]{1996ARA&A..34..155B}. Moreover, accurate models of galactic magnetic fields are crucial to inform the interpretation of many observables.

In this Letter, we demonstrate that differences in the treatments of cooling, star formation, and stellar feedback can yield very different magnetic field saturation values and morphologies. {\ksu We note that although a detailed analysis of magnetic amplification mechanism is beyond the scope of this paper, we do argue that the magnetic amplification history and saturation values in our simulations are reasonable.}
The remainder of this Letter is organized as follows: in \sref{S:methods}, we describe the initial conditions and investigated baryonic physics. In \sref{S:results}, we present the results. We summarize in \sref{s:discussion}.

\vspace{-0.5cm}
\section{Methodology} \label{S:methods}

Our simulations use {\sc gizmo} \citep{2015MNRAS.450...53H}\footnote{A public version of this code is available at \href{http://www.tapir.caltech.edu/~phopkins/Site/GIZMO.html}{\textit{http://www.tapir.caltech.edu/$\sim$phopkins/Site/GIZMO.html}}.}, a mesh-free, Lagrangian finite-volume Godunov-type code designed to combine the advantages of Eulerian and Lagrangian methods,
in its `meshless finite mass mode' (MFM). {\sc gizmo} is built on the gravity solver and domain decomposition algorithms of {\sc gadget-3} \citep{2005MNRAS.364.1105S}.
Magnetic fields are treated in the ideal-MHD limit. To eliminate spurious numerical divergence errors (i.e.~non-zero $\nabla\cdot B$), both the \cite{Dedner2002645} and
\cite{1999JCoPh.154..284P} divergence cleaning methods are applied. The details of the methods and tests are presented in \cite{2016MNRAS.455...51H} and \cite{2015arXiv150907877H}.

In this paper, two isolated (non-cosmological) galaxy models, MW and SMC, that have been used previously in various other works \citep{2011MNRAS.417..950H,2012MNRAS.421.3488H,2014MNRAS.442.1992H,2016arXiv160705274S} are studied. MW is a Milky Way-like galaxy with $(M_{\rm halo},M_{\rm bulge},M_{\rm disc},M_{\rm gas})=(210,2.1,6.8,1.3)\times 10^{10}$M$_\odot$, whereas SMC is Small Magellanic Cloud-mass dwarf with $(M_{\rm halo},M_{\rm bulge},M_{\rm disc},M_{\rm gas})=(290, 0.14, 1.9, 11)\times 10^8$ M$_\odot$. The gas mass resolution is $3500M_\odot$ for MW and $360M_\odot$ for SMC, which correspond to adaptive spatial resolutions of $\sim 50\,(n/{\rm cm^{-3}})^{-1/3}$ and  $\sim 20\,(n/{\rm cm^{-3}})^{-1/3}$ pc for the MW and SMC models, respectively. At these resolutions, the fastest-growing mode of amplification via the magnetorotational instability (MRI; $\sim 100$ pc) is well resolved once $B\sim \mu$G (it is not well-resolved initially).
 Detailed descriptions of the galaxy models can be found in \cite{2016arXiv160705274S}.
An initially uniform magnetic field with an amplitude of $0.01~\mu$G pointing in the positive-$z$ direction (i.e.~perpendicular to the initial disc galaxies) is assumed. We have confirmed that our results are insensitive to the initial magnetic field amplitude and orientation as long as the amplitude is sufficiently small that the magnetic pressure is initially dynamically unimportant. 
 
We simulate both galaxy models using the four different baryonic physics treatments summarized in \tref{tab:run}. The adiabatic runs include gravity and MHD, but radiative heating and cooling, star formation,
and stellar feedback are omitted. The no-feedback (`NoFB') runs include gravity, MHD, radiative heating and cooling, and star formation, but no stellar feedback. The FIRE runs add in explicit multi-channel
stellar feedback (from supernovae, stellar winds, photo-heating, and an approximate treatment of radiation pressure) from the Feedback in Realistic Environments (FIRE) project \citep{2014MNRAS.445..581H,Hopkins2017}. Specifically, the same version of code as the isolated runs in \cite{2016arXiv160705274S} (FIRE-1 feedback with MFM hydrodynamics) is used to facilitate comparison.
In both the NoFB and FIRE runs, stars form only from gas that is self-gravitating
at the resolution scale and has a density $n > 100$ cm$^{-3}$, and the instantaneous SFR density is assumed to be the molecular gas density divided by the local free-fall time (see \citealt{2014MNRAS.445..581H}). 
Except for the differences noted explicitly above, the code used for the runs is otherwise identical (and the NoFB and MHD runs specifically employ the exact same code as the corresponding
runs in \citealt{2016arXiv160705274S}). Finally, the `S\&H' runs employ the \citet[][hereafter SH03]{2003MNRAS.339..289S} model, which implicitly treats the ISM as gas with two phases (cold clouds \& ambient gas) in pressure equilibrium, but only has one explicit, single-phase gas with a barytropic equation of state (EOS) in the hydrodynamics. The model includes stochastic star formation via a Kennicutt-Schmidt-type prescription and implicitly treats the effects of supernova feedback by increasingly overpressurising the gas (relative to an ideal gas) with increasing density. To explore
potential self-consistent magnetically driven outflows, we do not employ the SH03 kinematically decoupled wind model. We have run the FIRE versions of the SMC and MW models at multiple resolutions spanning two orders of magnitude in mass resolution and confirmed that the global properties of the galaxies (e.g. SFHs and magnetic field strengths) are converged \citep{2016arXiv160705274S}. \ksu{We have also re-run both of the FIRE runs starting with an initial seed magnetic field strength an order of magnitude less ($10^{-3}~\mu$G) than our default value. Although the results differed during the initial exponential amplification phase, as expected, quantities such as the saturation values of the magnetic field strength in all and dense gas were similar.}

\begin{table}
\begin{center}
 \caption{Physics variations in our simulation suite}
 \label{tab:run}
 \begin{tabular}{cccc}
 \hline
\hline
Model         &Star Formation    &Cooling              & Feedback     \\
\hline
Adiabatic     &No    &None                 & None \\
NoFB          &Yes   &$10-10^{10}$ K       & None\\
FIRE          &Yes    &$10-10^{10}$ K       & FIRE  \\
S\&H          &Yes    &$10^4-10^{10}$ K     & Springel \& Hernquist  \\
\hline 
\hline
\end{tabular}
\end{center}
\end{table}

\vspace{-0.5cm}
\section{Results} \label{S:results}
\subsection{Magnetic field morphologies}

\begin{figure*}
\centering
\includegraphics[width=15.5cm]{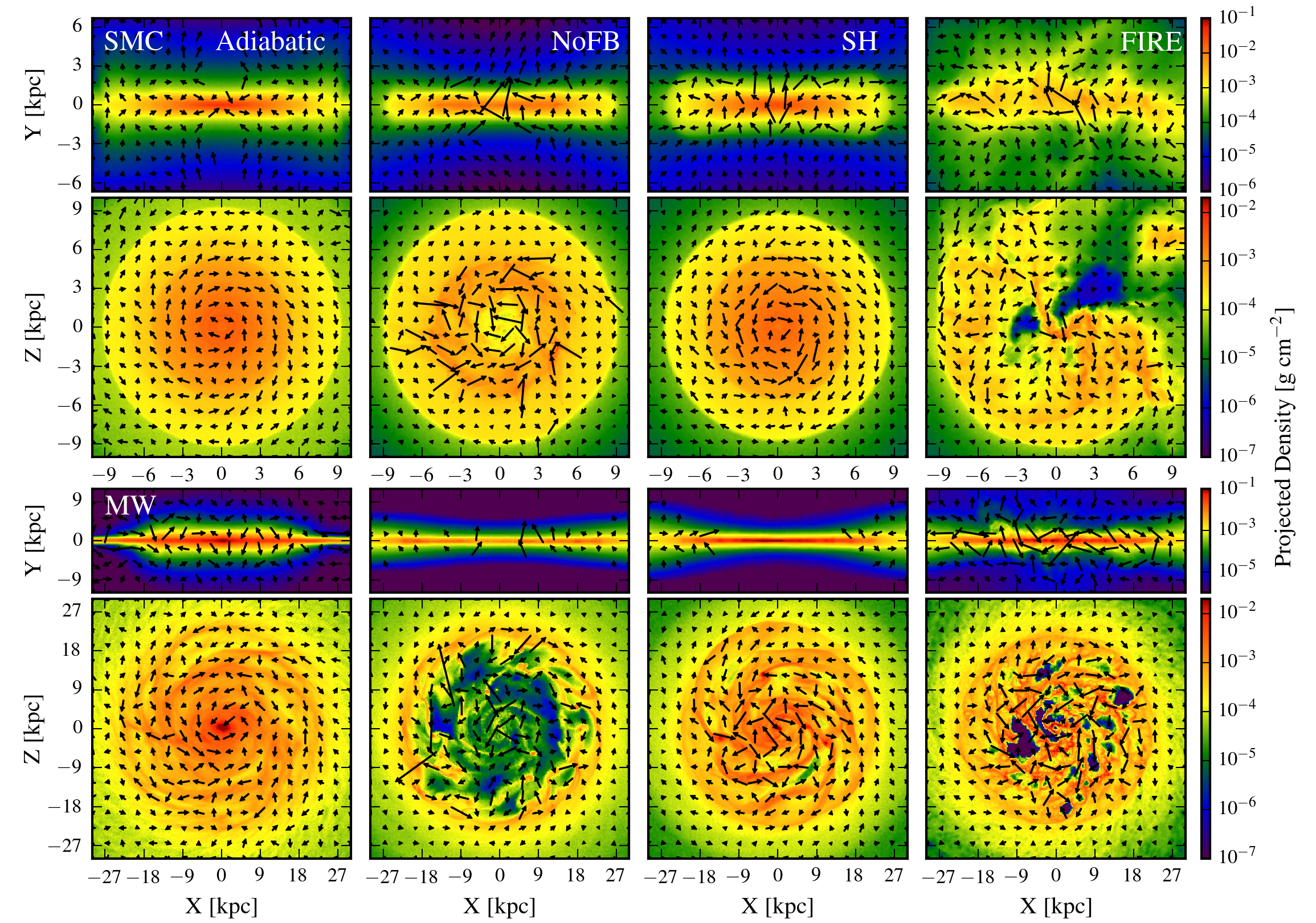}
\caption{Edge-on and face-on projections of the gas density of the simulated galaxies (MW, $t = 0.83$ Gyr: rows 1-2; SMC, $t = 0.69$ Gyr: rows 3-4).
Arrows indicate the relative magnitudes and directions of the magnetic field. Different columns correspond to different baryonic physics models. In all maps, the magnetic fields in dense clumps are not only stronger but also more randomly distributed.
No-feedback runs fragment most dramatically and therefore exhibit magnetic fields highly concentrated in dense clumps with random directions. Runs that employ the FIRE explicit stellar feedback model have irregular magnetic field distributions owing to supernova shocks, turbulence and outflows driven by stellar feedback, in addition to the greater fragmentation present in these runs compared with the Adiabatic and S\&H sub-grid stellar feedback runs. The latter two types of runs generally have smooth, highly ordered gas and magnetic field morphologies.}
\label{fig:mor_b}
\end{figure*}

\fref{fig:mor_b} shows edge-on and face-on projected gas density maps of the MW and SMC simulations. The MW runs are shown after
0.83 Gyr of evolution, whereas the SMC runs have been evolved for 0.69 Gyr.
The colour encodes the projected density, as specified by the colour bars on the right of each row (note that the scales in each row differ). The arrows in each subplot indicate the directions and relative magnitudes of the local magnetic field in a central slice with a thickness of 0.8 kpc for the SMC runs and 1.6 kpc for the MW runs. Note that the lengths of the arrows are separately rescaled in each subplot to cover a larger dynamic range. Different columns correspond to the different sets of included physics.

The gas and magnetic field morphologies of runs with different baryonic physics differ significantly. The local strength and orderedness of the magnetic field is strongly related to the local density, with higher gas density corresponding to stronger and more irregularly distributed local magnetic fields.  The NoFB runs fragment most dramatically because there is no stellar feedback to prevent catastrophic fragmentation. As a result, the magnetic fields are highly concentrated in the clumps and exhibit random directions. The magnetic fields in the FIRE runs are also highly irregular, not only because of fragmentation but also due to turbulence and outflows driven by strong stellar feedback \citep{2015MNRAS.454.2691M,2016arXiv161008523A,2015arXiv151005650H}. In contrast, the Adiabatic and S\&H runs have very smooth, well-ordered gas morphologies and magnetic fields.

\begin{figure}
\centering
\includegraphics[width=8.cm]{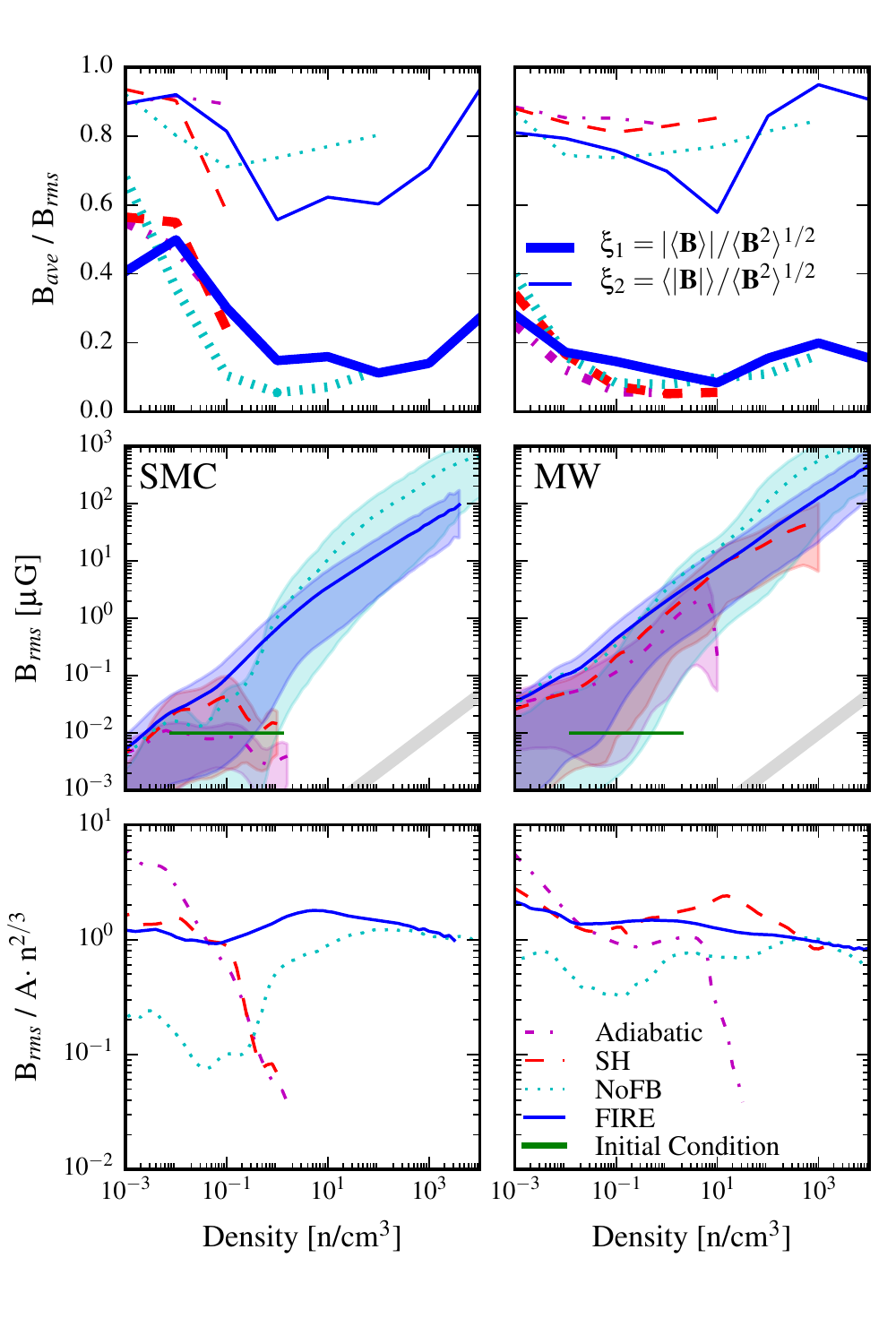}
\vspace{-1cm}
\caption{\emph{Top}: volume-weighted average values of $\xi_1\equiv|\langle{\bf B}\rangle|/B_{\rm rms}$ and $\xi_2\equiv\langle|{\bf B}|\rangle/B_{\rm rms}$ in 1 kpc cells in the galactic disc. $\xi_1$ and $\xi_2$ are both high in gas with density $<0.1 ~{\rm cm}^{-3}$ \ksu{averaged over 350-850 Myr for MW and 290-700 Myr for SMC}. In higher-density, fragmenting gas, the magnetic field is less spatially coherent, as indicated by the lower values of both $\xi_1$ and $\xi_2$ relative to the less-dense gas. In the density range $0.1-100 ~{\rm cm}^{-3}$, the FIRE runs generally have lower $\xi_2$ values, indicating a broader distribution of local magnetic field strengths. The S\&H and Adiabatic runs have much less gas with density $>0.1 ~{\rm cm}^{-3}$ than do the FIRE and NoFB runs. \emph{Middle}: relationship between the magnetic field strength and gas density of each resolution element \ksu{(time-averaged as in the top panel)}. The gray lines show a scaling of $n^{2/3}$ with arbitrary normalization. The shaded regions denote the 5th--95th percentile range of the magnetic field strength at each density. \emph{Bottom}: magnetic field strength divided by the best-fitting relation of the form $A\cdot n^{2/3}$ for each run.  In most density bins and most runs, $B_{\rm rms}$ scales as $n^{2/3}$, indicating isotropic flux freezing or equipartition between the magnetic and gravitational energies.
However, in the Adiabatic and S\&H runs, gas that is initially at `high' density cannot be compressed significantly owing to the high effective pressure; consequently, the field is not strongly amplified. }

\label{fig:Brho}
\end{figure}

To quantify the spatial coherence of the magnetic field, we calculate $\xi_1\equiv|\langle{\bf B}\rangle|/B_{\rm rms}$ and $\xi_2\equiv\langle|{\bf B}|\rangle/B_{\rm rms}$ in various regions of the disc. Low values of $\xi_1$ correspond to randomly directed local magnetic fields. Lower values of $\xi_2$ indicate broader local magnetic field strength distributions. In the top row of \fref{fig:Brho}, we plot $\xi_1$ and $\xi_2$ for gas at different densities. We first average (volume-weighted) over the gas particles from snapshots \ksu{(over the time period 350-850 Myr for MW and 290-700 Myr for SMC)} with similar densities within each 1 kpc cell and then plot the average of all cells in the disk region ($\Delta z=1$ kpc and $3<r<10$ kpc).\footnote{The center is excluded to prevent the average $\xi$ values from being dominated by extreme values.}

We generally find high $\xi_1$ and $\xi_2$ at densities $n\lesssim 0.1~{\rm cm}^{-3}$, indicating coherently directed magnetic field lines and narrowly distributed magnetic field strengths. In denser gas ($n \ga 0.1~{\rm cm}^{-3}$), fragmentation causes the local magnetic field to be amplified and become more randomly oriented, which lowers both $\xi$ values in the cell. The FIRE runs generally have the lowest $\xi_2$ over $0.1<n<100~{\rm cm}^{-3}$. This indicates a broader distribution of magnetic field strengths owing to disturbances from feedback. On the other hand, the S\&H and Adiabatic runs have almost no gas with $n>0.1~{\rm cm}^3$, which explains the extremely smooth magnetic field. At $n\gtrsim 100~{\rm cm}^{-3}$, $\xi_2$ increases again, indicating that the magnetic field in most gas within this density range has been amplified to a similar value.

\vspace{-0.6cm}

\subsection{Galactic magnetic field evolution}

\begin{figure}
\centering
\includegraphics[width=8.cm]{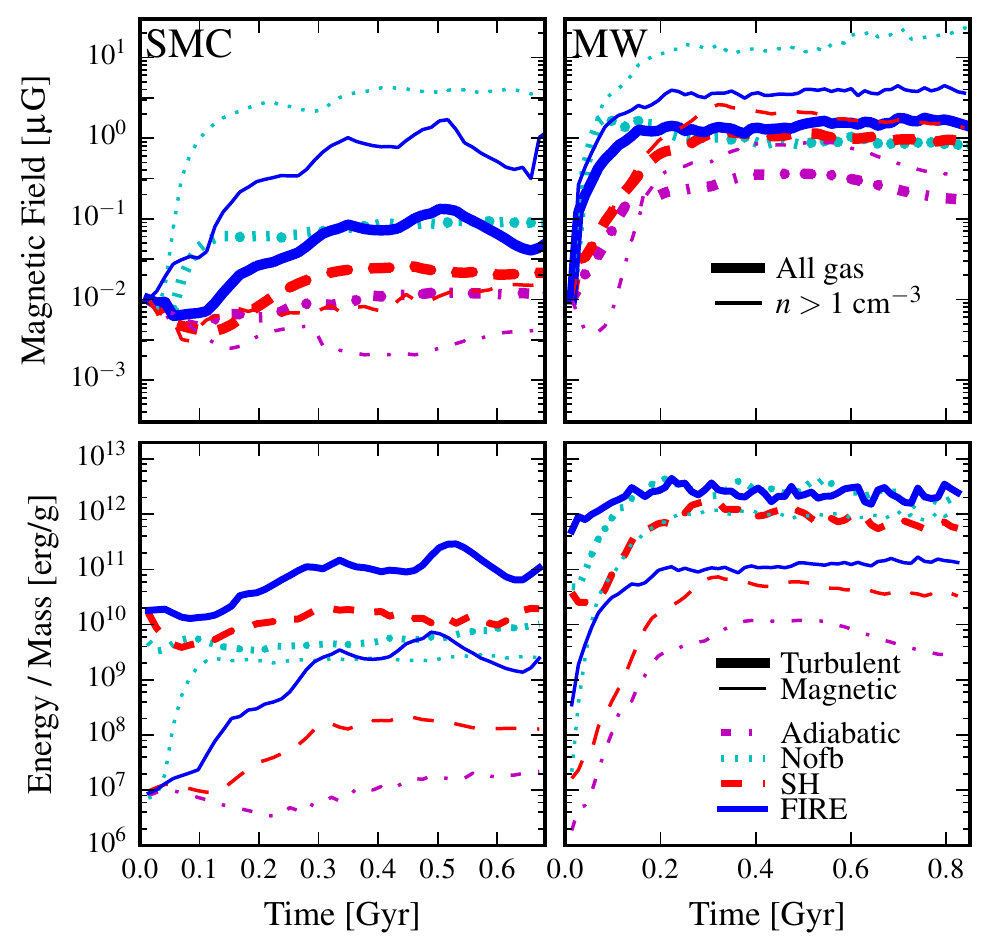}
\vspace{-0.5cm}
\caption{\emph{Top}: the rms magnetic field strength of gas in the disc
as a function of time. The different treatments of baryonic physics are indicated by the line color. Thick lines represent the rms magnetic field strength computed considering all gas, $B_{\rm rms}^{\rm all}$, whereas thin lines indicate the results for gas with density $>1$ cm$^{-3}$, $B_{\rm rms}^{\rm dense}$. Greater gas fragmentation is associated with a higher magnetic field strength saturation value. The Adiabatic and S\&H runs, which do not fragment as much as the no-feedback and FIRE runs, exhibit lower saturation values. \ksu{\emph{Bottom}: comparison of turbulent and magnetic energy per unit mass as a function of time. In the FIRE and S\&H runs, the magnetic energy saturates to $\sim2-6\%$ of the turbulent energy, consistent with the idealized supersonic turbulent dynamo.} 
}
\label{fig:Brms}
\end{figure}

\fref{fig:Brms} shows the evolution of the rms magnetic field strength of the gas in the disc (defined as a cylinder of height 2 kpc and radius 10 kpc).  We compare the rms $B$ in all gas and just gas with $n\gtrsim 1 ~{\rm cm}^{-3}$. \ksu{In the MW runs, the magnetic field exponentially grows from the initial value ($10^{-2}~\mu$G throughout the volume) at the start of simulation. Even in the MW Adiabatic run, the magnetic is amplified significantly through gravitational compression. In the SMC runs, on the other hand, the saturation values for all but the very dense particles are approximately equal to the initial value of magnetic field ($10^{-2}~\mu$G); hence, the amplification is much weaker for the SMC runs, especially the Adiabatic run. In the FIRE-low runs, the amplification is more obvious. Regardless of the initial magnetic field strength, the saturation value for each IC (using the FIRE baryonic physics model) is robust.}

Since the denser gas has much stronger magnetic fields, the time evolution and saturation value of the field is closely tied to the amount of fragmentation in each run.  The Adiabatic runs generally fragment the least and therefore have the lowest magnetic field strength. In the S\&H runs, the stiff EOS adopted in the model and the temperature floor ($10^4$~K) prevent gas from being compressed and fragmenting as much as in the NoFB and FIRE runs. As a result, the magnetic fields in the S\&H runs are also weaker overall. However, in gas with densities less than or similar to the galactic average, the S\&H model yields quite similar field strengths to the NoFB or FIRE runs. Due to the absence of feedback, the NoFB runs fragment most severely, and the densest gas therein has the highest rms magnetic field strengths.  

The magnetic field strength in the gas disc in the MW FIRE run saturates at $\sim 2~\mu$G, and the field strength of ``dense''  ($n>1 \,{\rm cm}^{-3}$) gas saturates at $\sim 5~\mu$G. These values are consistent with observations \citep{1989ApJ...343..760R,1996ARA&A..34..155B,2006ApJ...642..868H,2007EAS....23...19B,2008RPPh...71d6901K}. At higher densities ($n\sim 100~{\rm cm}^{-3}$), comparable to \ion{H}{i} clouds in the cold neutral medium or molecular clouds, $|{\bf B}|$ is typically a few 10s of $\mu$G, again similar to observations.

\ksu{As shown in the second row of \fref{fig:Brms}, the} magnetic energies in the FIRE and S\&H runs saturate to $\sim2-6\%$ of the turbulent energy \ksu{\footnote{The turbulent energy calculation is performed using the method described in \cite{2016arXiv160705274S}. This is not applicable for Adiabatic runs, so these runs are not included here.}. This is true regardless of the initial magnetic field strength}, consistent with \ksu{theoretical predictions \citep{2015PhRvE..92b3010S} and idealized simulations of the supersonic turbulent dynamo \citep{2011PhRvL.107k4504F,2014ApJ...797L..19F,2016MNRAS.461.1260T} and  other galaxy simulations with feedback \citep{2010ApJ...716.1438K,2010A&A...523A..72D,2016arXiv160705274S,2017arXiv170405845R}. }\ksu{We therefore believe that our magnetic amplification histories and saturation values are reasonable.} In contrast, in the NoFB runs, the magnetic energy reaches $\gtrsim 30\%$ of the turbulent energy, \ksu{indicating equipartition between the turbulent and magnetic energy}. 

\ksu{As can be seen from the evolution of the magnetic field strength and  energy, the exponential growth happens during the first $\sim 0.2-0.3$ Myr, which is shorter than the time scale for amplification driven by galactic global motion. This indicates that either local turbulent motion or local gravitational motion dominates the magnetic amplification during this time. However, separating these two processes is difficult in galactic simulations, since the local gravitational and turbulent energy can be in equipartition. A detailed study of the dominant amplification mechanism(s) will be presented in a future work.}

\vspace{-0.5cm}
\subsection{Relationship between magnetic field strength and density}

\ksu{The second row of} \fref{fig:Brho} shows the relationship between the local magnetic field strength and local gas density. Over most of the density range, $B_{\rm rms}\propto n^{2/3}$. In low-density gas, this is a result of isotropic flux freezing in gas compression dominating the field amplification (or in expansion, for the lowest-density gas). In denser gas, this indicates either isotropic flux freezing or equipartition between the magnetic and gravitational energies in collapsed structures with constant mass. The latter holds because
\begin{align}
\frac{{\bf B}^2}{4\pi} V\sim \frac{GM^2}{r} {\bf \rightarrow} B\propto \frac{1}{(rV)^{1/2}}\propto \frac{1}{V^{2/3}}\propto n^{2/3}.
\end{align}

\ksu{In the MW runs, gas in the initial disc (green line in \fref{fig:Brho}) has a magnetic field strength at least 1 order of magnitude less than the average magnetic field in later snapshots and outside the shaded region spanned by the gas particles therein. In the SMC runs with the fiducial seed field strength, on the other hand, the magnetic field strength is relatively high to begin with, so the initial magnetic field strength is marginally within the shaded region, indicating weaker amplification. However, in the `FIRE-low' runs for both model galaxies, although the initial magnetic field strength is one order of magnitude lower, the $B_{\rm rms}-n$ curves from the later snapshots are almost identical to the curves in the default-seed-value `FIRE' runs.  If the same plots are made for individual time snapshots, $B_{\rm rms}\propto n^{2/3}$ holds at all times, albeit with systematic shifts in the normalisation. Within the first $\sim 0.2-0.3$ Myr, the lines move toward the upper left and approach the lines shown in \fref{fig:Brho}. This indicates that although isotropic flux freezing or equipartition between the magnetic and gravitational energies dominates the field amplification as gas undergoes density changes, other amplification mechanisms (local turbulent or gravitationally driven amplification) occur in these runs, especially within the exponential growth phase of the magnetic fields.}

Gas at the high-density ends of both the Adiabatic and S\&H runs deviates from the adiabatic curve (i.e.~is less amplified than expected). This is an artifact caused by the fact that gas initially
at `high' density in those runs cannot be significantly compressed further. For the Adiabatic runs, the reason is that the lack of cooling results in high pressure in high-density regions. In the S\&H run, the lack of cooling below $10^4$ K and the effective EOS have the same effect. In particular, owing to the small depth of the potential well of the SMC model, gas in the Adiabatic and S\&H runs is only weakly compressed; thus, the magnetic field is not significantly amplified. 

\vspace{-0.7cm}
\section{Summary and discussion} \label{s:discussion}

We have demonstrated that the morphology and saturation value of galactic magnetic fields strongly depend on baryonic physics, specifically cooling, star formation and stellar feedback. This is primarily because baryonic physics affect the magnetic field strength through altering the amount of fragmentation and balance amongst different ISM phases.
Moreover, strong turbulence and outflows driven by multi-channel stellar feedback  (present in our FIRE runs) can further increase the amount of randomness in the magnetic field morphology.
The Adiabatic (no cooling, no star formation) and S\&H (``effective equation of state'', unresolved ISM) runs have significantly lower magnetic field saturation values in dense gas compared with the NoFB (cooling and star formation but no feedback) and FIRE (cooling, star formation, and feedback) runs, and the former tend to have
more ordered large-scale magnetic fields. It appears that the S\&H model (which is used by many modern galaxy formation simulations, either directly or with modifications, such as IllustrisTNG, \citealt{2017arXiv170302970P}; EAGLE, \citealt{2015MNRAS.446..521S}; and MUFASA, \citealt{2016MNRAS.462.3265D}) works reasonably well for gas with density lower than or equal to the galactic average but suppresses gas fragmentation in higher-density gas, thus causing discrepancies in the magnetic field amplification. 

We find that in the simulated galactic discs, the field strength scales with density as $B\propto n^{2/3}$, similar to the results for galactic cores found by \citet{2017arXiv170107028P}. This results from a combination of isotropic flux freezing in compression/expansion without a preferred direction and equipartition between the magnetic and gravitational energies (in dense, self-gravitating gas). \ksu{ We caution that although the aforementioned processes dominate the field amplification as gas undergoes density changes, they are not the only amplification mechanisms at work.}

Note that although Zeeman observations \citep{2010ApJ...725..466C} suggest $B\propto n^{0}$ at low density and $B\propto n^{2/3}$ at high density, the scatter in the data points is quite large. Moreover, the magnetic field strength inferred is highly correlated with whether the object is an \ion{H}{I} cloud/dark cloud or a molecular cloud, and the scaling between the magnetic field strength and density is weak within each category. Since both of these categories of objects have nearly constant surface density \citep{1986A&AS...66..505W,1994AJ....107.1003C,2002ApJ...569..157W,2004ApJ...612L..29B}, the aforementioned observations might actually constrain the scaling between magnetic field strength and surface density (instead of volume density) and are not necessarily inconsistent with our results.

\citet{RiederTeyssier2016,2017arXiv170405845R} have also argued that stellar feedback is crucial for magnetic field amplification; in  simulations of isolated gas cooling haloes, they find that supernova feedback drives a turbulent dynamo that is the dominant source of magnetic field amplification. A detailed comparison of our results is beyond the scope of this work, but it is worth noting that we both agree that the effects of stellar feedback must be captured accurately in order to model the amplification of galactic magnetic fields. 

Our results also have important implications for incorporating cosmic rays, which may play an important role in driving galactic outflows \citep[e.g.][]{Uhlig2012,Hanasz2013,Booth2013,Salem2014a}, because cosmic rays propagate along magnetic field lines. We have shown that `effective' treatments of stellar feedback lead to significantly more ordered magnetic fields compared with simulations with explicit stellar feedback; it is thus likely that in the former types of simulations, cosmic rays will propagate
over large scales and drive outflows more easily than in simulations with explicit stellar feedback. 



\vspace{-0.8cm}
\acknowledgments
The Flatiron Institute is supported by the Simons Foundation. Support for PFH was provided by an Alfred P.~Sloan Research Fellowship, NASA ATP Grant NNX14AH35G, and NSF Collaborative Research Grant \#1411920 and CAREER grant \#1455342. CAFG was supported by NSF through grants AST-1412836 and AST-1517491, and by NASA through grant NNX15AB22G.
DK was supported by NSF grant AST-1412153 and a Cottrell Scholar Award from the Research Corporation for Science Advancement. Numerical calculations were run on the Caltech compute cluster ``Zwicky'' (NSF MRI award \#PHY-0960291) and allocation TG-AST130039 granted by the Extreme Science and Engineering Discovery Environment (XSEDE) supported by the NSF. \\

\bibliographystyle{mnras}
\bibliography{mybibs}


\label{lastpage}

\end{document}